\RequirePackage{fix-cm}
\documentclass[]{svjour3}                     
\smartqed  
\usepackage{graphicx}
\usepackage{amsmath}

\usepackage{color,soul}

\begin{document}

\title{Computationally efficient method for retrieval of atmospherically distorted astronomical images
}
\subtitle{}


\author{Arun Surya     \&
        Swapan K. Saha 
}


\institute{A.Surya \and S.K.Saha ( Retired )\at
              Indian Institute of Astrophysics \\
               2nd Block Koramangala\\
               Bangalore                         
              \email{arun@iiap.res.in}}

\date{Received: date / Accepted: date}

\maketitle

\begin{abstract}
Speckle Imaging based on triple correlation is a very efficient image reconstruction technique which is used to retrieve Fourier phase information of the object in presence of atmospheric turbulence. { We have developed both Direct Bispectrum and Radon transform based Tomographic speckle masking algorithms to retrieve atmospherically distorted astronomical images.} The latter is a much computationally efficient technique because it works with one dimensional image projections. Tomographic speckle imaging provides good image recovery like direct bispectrum but with a large improvement in computational time and memory requirements. { The algorithms were compared with speckle simulations of aperture masking interferometry with 17 sub-apertures using different objects.} The results of the computationally efficient tomographic technique with laboratory and real astronomical speckle images are also  discussed.

\end{abstract}

\section{Introduction}
Speckle interferometry \cite{lab70} is a method that is used to reconstruct diffraction limited spatial Fourier spectrum and image features of astronomical
objects by overcoming the image distortion caused by the atmospheric turbulence. Used together with aperture masking techniques (speckle techniques with non
redundant pupils) it has already made impacts in several important fields in astrophysics, mainly in the field of origin and evolution of stellar systems.{ The advantages of such  techniques are described in depth by Saha} (\cite{sks1,sks2,sks3}). {However, classical speckle interferometry falls short of obtaining  the phase information of the object and only provides a second-order moment (power spectrum) analysis which gives the amplitude of the object Fourier transform.} Triple Correlation technique and other advanced image retrieval methods have been 
developed to allow the reconstruction of the Fourier phase information. Such algorithmic techniques retrieve diffraction limited information from the short
exposure images of the object. The triple correlation technique developed by Weigelt and Lohmann (\cite{wei77,loh83}) is a third-order moment (bispectrum) analysis which yields the Fourier phase of the image, allowing the complete image retrieval. The advantages of this technique is in providing  information about the object phases with better signal to noise ratio (SNR). The disadvantage of this technique is that it demands very large computational resources with 2-dimensional data since the calculations are 4-dimensional. It requires
extensive evaluation-time and data storage requirements, even if the correlations are performed by using digitized images on a computer. Tomographic methods using
Radon transform offers a better alternative since they are computationally efficient. {We have developed triple correlation algorithms based on both direct bispectrum and radon transform to process the speckle frames of the night time astronomical objects.} The results obtained from the developed code with numerical simulations, laboratory simulations of aperture masking interferometry and real speckle images of binary stars  are presented in this paper.

\section{Tomographic Speckle Imaging}

Bispectrum technique is a very powerful technique which allows to recover  astronomical images
from several short exposure speckle images. Since the bispectrum for 2-dimensional images are 4-dimensional, the processing time and memory requirements are huge for large images. We have developed a tomographic technique which reduces the memory and processing requirements in bispectrum technique. The tomographic technique transforms the images in to a set of projections
and applies bispectrum based reconstruction on these one dimensional projections. {By transforming 2-dimensional speckle images in to its 1-dimensional speckle projections,
huge savings in memory and processing time have been achieved.} The developed algorithm combines two powerful mathematical techniques, Triple Correlation and Radon Transformation.

\subsection{Triple Correlation}
The triple correlation technique developed by Lohmann et al \cite{loh83} runs as 
follows. The object speckle pattern, ${\ I}({\bf x})$, is 
multiplied with an appropriately shifted version of it, i.e., ${  I}({\bf x} + {\bf x}_1)$. 
The result is then correlated with ${I}({\bf x})$. 

$${  I^{(3)}}{\bf (x}_1, {\bf x}_2) = \langle\int^{+\infty}_{-\infty}{  I}
{\bf (x)}{  I}({\bf
x} + {\bf x}_1) {  I}({\bf x} + {\bf x}_2)d{\bf x}\rangle \eqno(1)$$

\noindent
where, ${\bf x}_j = {\bf x}_{jx} + {\bf x}_{jy}$ is the 2-dimensional spatial 
co-ordinate vector. $\langle \rangle$ stands for ensemble average.
\vspace{0.2cm}

The Fourier transform of the triple correlation is called bispectrum and its
ensemble average is given by,

$$\langle\widehat{  I}^{(3)}({\bf u}_1, {\bf u}_2)\rangle = \langle\widehat{  I}({\bf u}_1) 
\widehat{  I}^\ast 
({\bf u}_1 + {\bf u}_2) \widehat{  I}({\bf u}_2)\rangle \eqno(2)$$

\noindent
where
$\widehat{  I}({\bf u}) =\int{  I}({\bf x})e^{-i2\pi{\bf u}.{\bf x}} 
d{\bf x}, \widehat{  I}^\ast({\bf u}_1 + {\bf u}_2) = \int{  I}({\bf x})
e^{i2\pi({\bf u}_1 + {\bf u}_2).{\bf x}}d{\bf x}, 
{\bf u}_j = {\bf u}_{jx} + {\bf u}_{jy}. $ 
\vspace{0.2cm}

In the second order moment phase of the object's Fourier transform is lost,
but in the third order moment (bispectrum) it is preserved. The argument
of equation (2) can be expressed as,

$$arg[\widehat{  I}^{(3)}({\bf u}_1, {\bf u}_2)] = 
\phi_b({\bf u}_1, {\bf u}_2) = \phi({\bf u}_1) - \phi({\bf u}_1 + {\bf u}_2) + 
\phi({\bf u}_2) \eqno(3)$$

Equation (3) gives the 
phase of the bispectrum. Observed image is the convolution of the object
and the point spread function (PSF) of the combination of telescope and the atmosphere.
Its Fourier transform is the product of the Fourier transform of the object
and the transfer function of the telescope and the atmosphere.

$$\widehat{  I} = \widehat{  O}({\bf u})\cdot\widehat{  S}({\bf u}) \eqno(4) $$

\noindent
Invoking equation (4) into equation (2),

$$<\widehat{  I}^{(3)}({\bf u}_1, {\bf u}_2)> = \widehat{  O}({\bf u}_1)\widehat{  O}^
\ast({\bf u}_1+{\bf u}_2)\widehat{  O}({\bf u}_2) <\widehat{  S}({\bf u}_1)
\widehat{  S}^\ast({\bf u}_1+{\bf u}_2)\widehat{  S}({\bf u}_2)> \eqno(5)$$

Thus the image bispectrum is the product of object bispectrum and bispectrum transfer function 
$<\widehat{  S}({\bf u}_1)\widehat{  S}
^\ast({\bf u}_1 + {\bf u}_2)\widehat{  S}({\bf u}_2)>$.  
It has been proved elsewhere \cite{loh83}  that this transfer function is a real-valued function. {Thus the phase  of the averaged image bispectrum is equal to the phase of the object bispectrum.} This allows for the opportunity to extract real phase information of the object from the 
object bispectrum. The modulus $\mid\widehat{  O}({\bf u})\mid$ and phase $\phi({\bf u})$ of the 
object Fourier transform $\widehat{  O}({\bf u})$ can be derived from the object 
bispectrum $\widehat{  I}_{  O}^{(3)}({\bf u}_1, {\bf u}_2)$.  
{The object phase-spectrum is thus encoded in the term
$e^{i\phi_b({\bf u}_1, {\bf u}_2)} = e^{i[\phi({\bf u}_1) 
- \phi({\bf u}_1 + {\bf u}_2) + \phi({\bf u}_2)]}$.}

{Equation (3) is a recursive equation for the phase of the object 
bispectrum at coordinate ${\bf u} = {\bf u}_1 + {\bf u}_2$.} 
{The phase of the
object Fourier transform  at $({\bf u}_1
+ {\bf u}_2)$ can be expressed as,}

$$\phi({\bf u}_1 + {\bf u}_2) = \phi({\bf u})  
= \phi({\bf u}_1) + \phi({\bf u}_2) - \phi_b({\bf u}_1, {\bf u}_2) \eqno(6)$$

If the object spectrum at ${\bf u}_1$ and ${\bf u}_2$ are known,
the object phase-spectrum at $({\bf u}_1 + {\bf u}_2)$ can be computed. {Thus Triple correlation technique provides a very good method of retrieving the Fourier phase of the object distribution. Modified
version of the same technique is also applied to extract information from closure of phase measurements in long baseline optical interferometry.}

\setcounter{equation}{7}
\subsection{Radon Transform}

\begin{figure}
\center
\includegraphics[width=100mm]{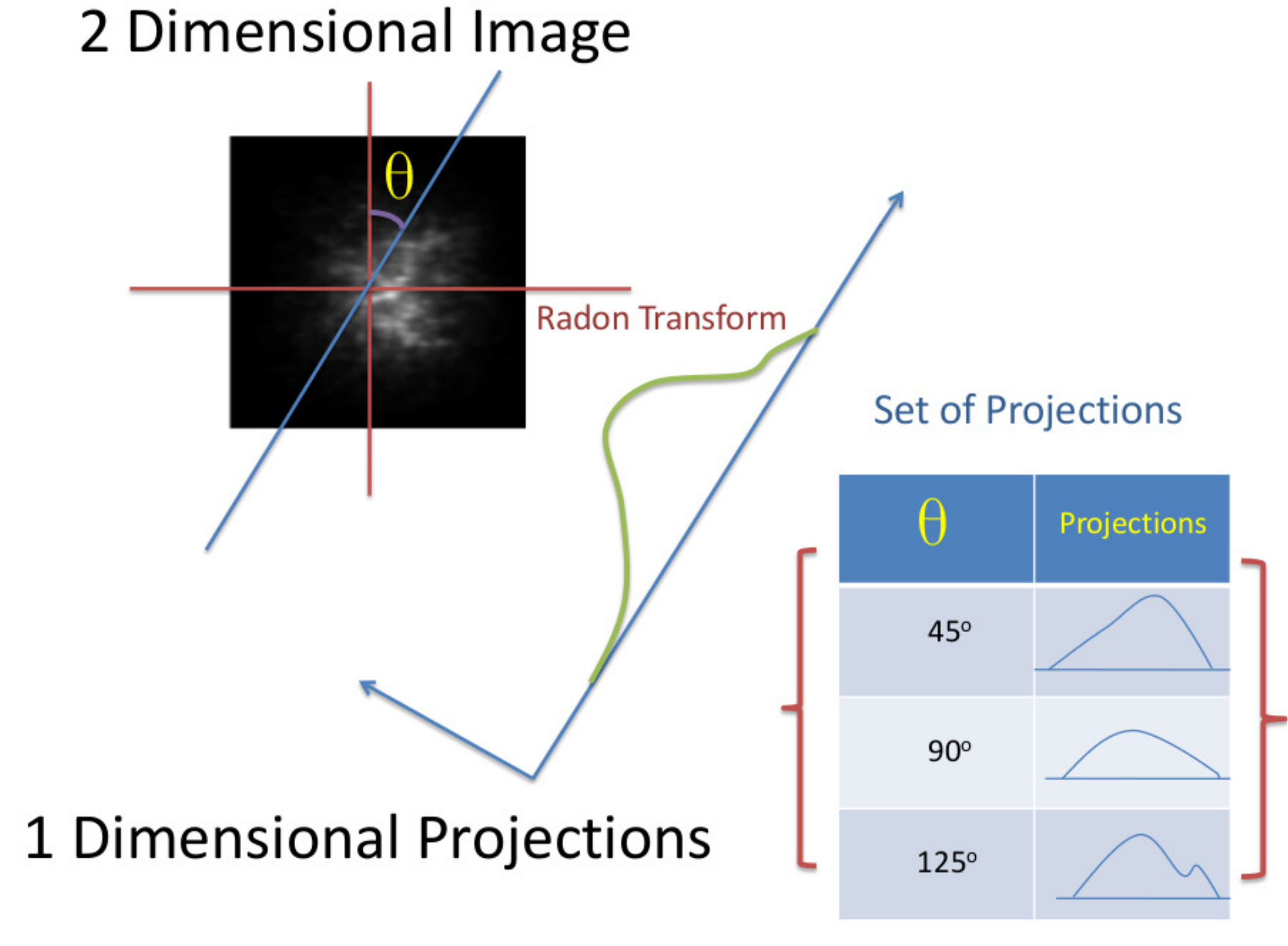}
\caption{ A 2-dimensionsl speckle image and its sample 1-dimensional projections obtained at different angles using Radon Transform.
}

\label{fig:1}       
\end{figure}
{Radon Transform is a method by which a  2-dimensional image is transformed in to a set of 1-dimensional projections. This transformation gives us the capability to handle a 2-dimensionalsignal as a set of one dimensional
signals.} Radon transform for function $f(x,y)$ is described mathematically as

\begin{equation}
{\theta(x')=\int_{-\infty}^{+\infty}{f(x'cos\theta - y'sin\theta + x'sin\theta - y'cos\theta) dy'}}
\end{equation}

\begin{equation}
 \begin{bmatrix}
       x'   \\[0.3em]
       y'   
     \end{bmatrix} = \begin{bmatrix}
       cos\theta & sin\theta  \\[0.3em]
       -sin\theta & cos\theta
     \end{bmatrix} \begin{bmatrix}
       {x}   \\[0.3em]
       {y}   
     \end{bmatrix}
\end{equation}

In radon transform based triple correlation method, we will radon transform each frame to a set of projections {$I_1,I_2,..I_n$}. Then find the averaged Triple Correlation for each projection over all frames 

\begin{equation}
 <I^{(3)}(x_1,x_2)>=\langle\int_{-\infty}^{+\infty} I(x)\cdot I(x+x_1)\cdot I(x+x_2) dx \rangle
\end{equation}

or  compute the averaged bispectrum
\begin{equation}
 <\hat{I}^{(3)}(u_1,u_2)>=\langle \hat{I}(u_1) \cdot \hat{I}(u_2) \cdot \hat{I}^\star(u_1+u_2)\rangle
\end{equation}
From this averaged bispectrum the real projections are reconstructed and they are inverse radon transformed to recover the original signal. Radon transform based image retrieval takes less computational time and also requires less memory at the expense of reconstruction quality.

\section{Algorithms}
\subsection{Direct Bispectrum}

Using triple correlation based recursive technique we have developed a code in MATLAB to process images degraded by atmospheric turbulence. The code works on 2-dimensional images and uses its corresponding  4-dimensional bispectrum 
to recover the Fourier phase information of the object. The direct bispectrum code can process 200 x 200 pixels images of 300 frames in 15 minutes in a Core 2 Duo Intel computer with 4 GB of RAM. The unit amplitude phasor method,
used in algorithms by Sridharan \cite{sri00}, is applied in the code for phase reconstruction. The code uses direct computation of  the 4-dimensional bispectrum $I^{(3)}(u,v,u^{'},v^{'})$ which demands large amount of computer memory. The 4-dimensional bispectrum is computed and averaged out for all the speckle frames. The retrieved Fourier phase from bispectrum is combined with Fourier amplitude from speckle interferometry to reconstruct the object.

\subsection{TSM Algorithm}

In the TSM algorithm we combine triple correlation and radon transform techniques to develop a robust and fast algorithm to retrieve images. The steps in the developed algorithm is as follows
\begin{enumerate}
  \item Radon Transform each 2-dimensional frame in to set of projections $I_1,I_2,I_3..etc$
  \item Find the averaged bispectrum $\langle\hat{I}^{(3)}(u_1,u_2)\rangle$ for each projection over all frames. 
  \item Retrieve the Fourier phase $\phi({u})$ of the object projection from the averaged bispectrum using recursion formula used in equation (6)  .
  \item Retrieve the Fourier amplitude of the object projection from the averaged power spectrum $\langle\hat{I}^{(2)}(u_1,u_2)\rangle$.
  \item Combine the Fourier phase and amplitude to recreate the true object projections.
  \item Inverse radon transform the retrieved projections to obtain the object image.
\end{enumerate}

\section{Numerical Simulations}

\begin{figure}
\center
  \includegraphics[width=60mm]{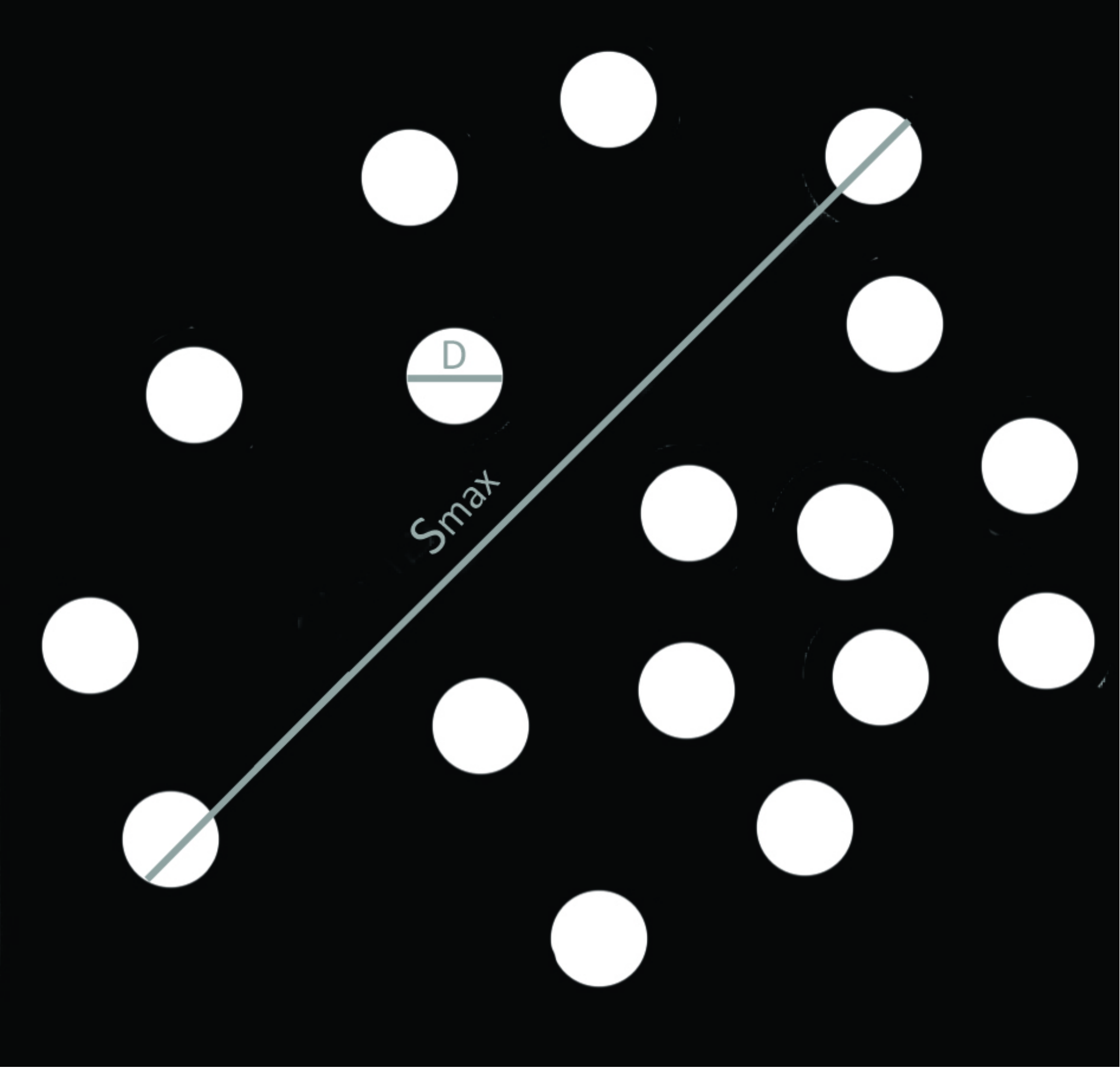}
\caption{The aperture used for the simulation}

\label{fig:2}       
\end{figure}

\begin{figure}
\center
  \includegraphics[width=80mm]{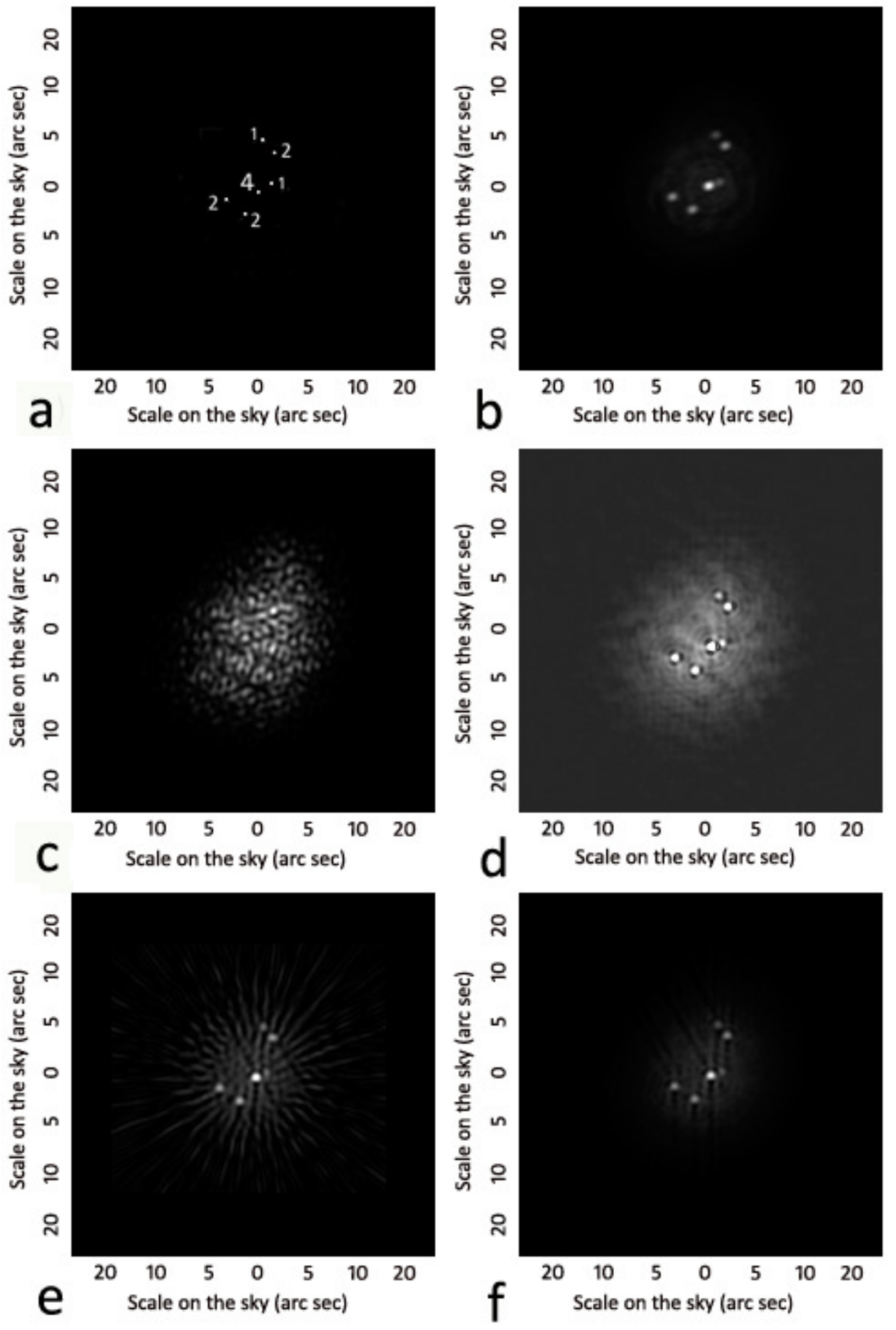}
\caption{The image recovery results from the tomographic speckle imaging algorithm of a 6 star group using a 17 hole aperture mask. The aperture consists of randomly arranged 10 cm holes inside 1 m radius disk. a) The 6 star object distribution used in the simulation with the corresponding brightness ratio. b) The cophased image from the aperture masked mirror. c) The simulated speckle image of the object by imaging through turbulence. d) The recovered image using Direct Bispectrum with 100 frames. e) The recovered image using Tomographic speckle imaging from 18 projections with 100 frames. f) The recovered image using Tomographic speckle imaging from 180 projections with 100 frames.
}

\label{fig:3}       
\end{figure}

\begin{figure}
\center
  \includegraphics[width=80mm]{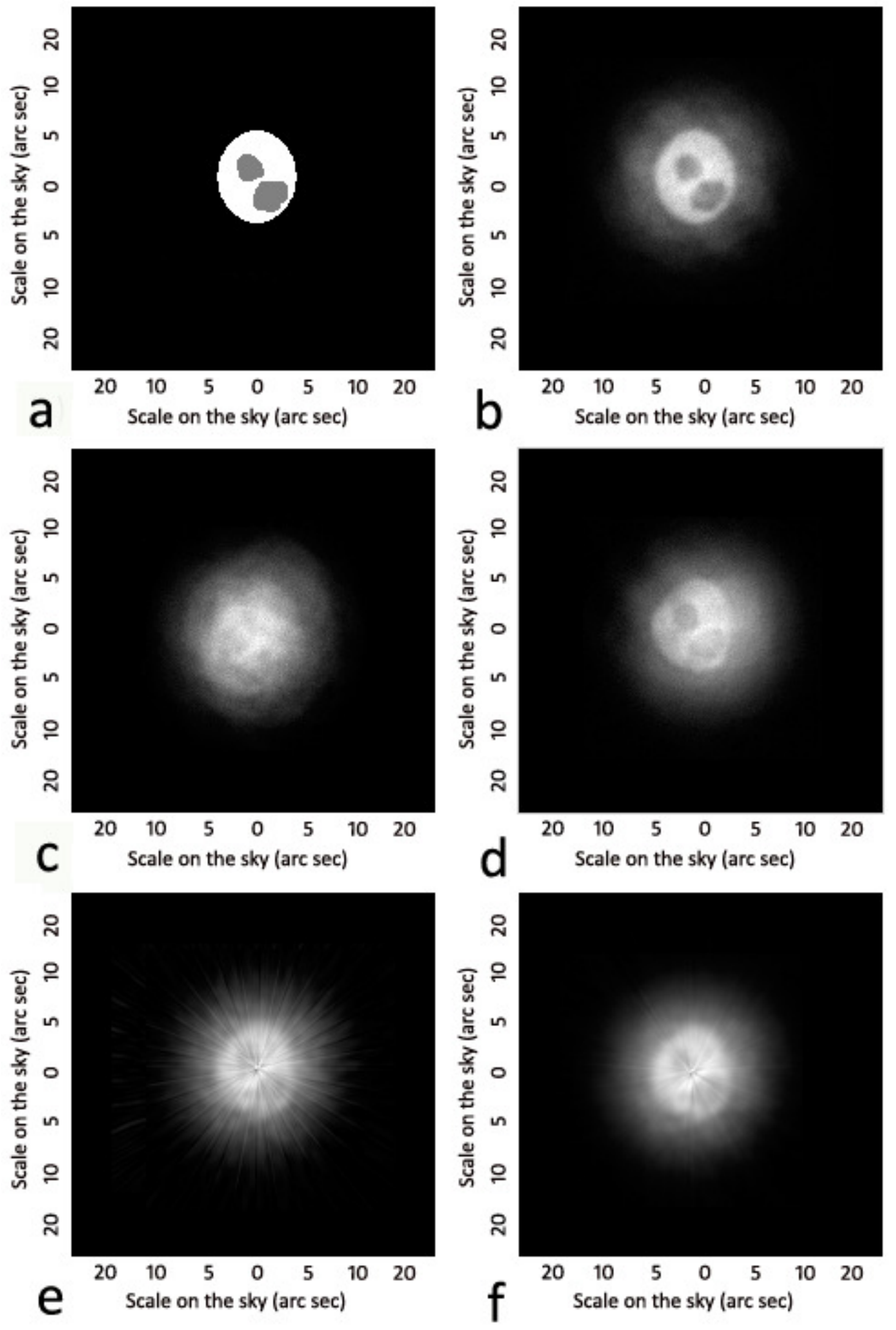}
\caption{The image recovery results from the tomographic speckle imaging algorithm of an extended planet object using a 17 hole aperture mask. The aperture consists of randomly arranged 10 cm holes inside 1 m radius disk. a) The extended object used in the simulation with the corresponding brightness ratio. b) The cophased image from the aperture masked mirror. c) The simulated speckle image of the object by imaging through turbulence. d) The recovered image using Direct Bispectrum with 100 frames. e) The recovered image using Tomographic speckle imaging from 18 projections with 100 frames. f) The recovered image using Tomographic speckle imaging from 180 projections with 100 frames.
}

\label{fig:4}       
\end{figure}

Aperture masking is a technique of using masked apertures on a single aperture telescope \cite{bald86,Chris87} and to process the fringe systems thus obtained to reconstruct the image. Aperture masking has been used together with speckle imaging to obtain results with better SNR \cite{sks4}. In our study we have carried out laboratory and numerical simulations of speckle imaging with aperture masks having randomly arranged subapertures.
The numerical simulations were used to compare reconstructions achieved by direct bispectrum and tomographic speckle masking. In these simulations we use a mask with non redundant aperture of 17 holes. The aperture is shown in figure \ref{fig:2}.
The simulation considers the imaging PSF to be
\begin{equation}
{I(x,y)=  A(x,y) \cdot \left| \sum_{j=1}^{N}    e^{-\frac{2\pi i}{\lambda}(xu_{j}+yv_{j})} \cdot e^{i{\phi_j}}\right|^{2}}
\end{equation}

with $u_j$ and $v_j$ the positions of the $j^{th}$ subaperture, $\phi_j$ the corresponding atmospheric piston error and with $A(x,y)$ being the diffraction function corresponding to subaperture give by,

\begin{table}
\center\begin{tabular}{|l|c|c|}
\hline 

 &   \tabularnewline

Frames used in speckle masking & 100 \tabularnewline
Plate scale for each frame & 0.2 arc sec /pixel \tabularnewline
Wavelength & 500 nm \tabularnewline
Fried Parameter  & 0.2 m \tabularnewline
Wind Velocity     & 10 m/s \tabularnewline
Maximum Baseline & 1 m \tabularnewline
Minimum Baseline (between subapertures)& 20 cm \tabularnewline
Diameter of subaperture & 10 cm \tabularnewline

\hline
\end{tabular}

\caption{Parameters used in the numerical simulations of aperture masking speckle imaging}
\label{tab1}
\end{table}

\begin{equation}
A(x,y)=\left|\frac{2 J_1(\rho R)}{\rho R}\right|^2.
\end{equation}

where $\rho = \frac{2\pi }{\lambda} \sqrt{x^2+y^2}$ is the radial distance in reciprocal space and $R$ is the radius of subaperture. For the numerical simulations we considered the maximum baseline of the mask $S_{max}$ to be 1 m and the minimum baseline between the subapertures as 20cm. The diameter of the subaperture $D$ is 10 cm.
The atmospheric piston errors of subapertures were taken from a Kolmogorov phase screen of Fried parameter 20 cm which was been simulated using the Fast Fourier Transform based power density method \cite{rglane}. The phase screens were moved with a wind velocity of 10 m/s between successive
speckle frames to simulate seeing changes in real conditions.  All the major parameters of the simulation are shown in table \ref{tab1}. The figures  \ref{fig:3} and \ref{fig:4} shows the reconstruction results obtained with tomographic speckle imaging reconstruction of simulated aperture masking images. Figure \ref{fig:3} shows the results with a binary star and figure \ref{fig:4} shows the reconstruction results with
an extended object. 
{ We have quantified the reconstruction quality using 2-dimensional correlation coefficient $c$ which measures the correlation between the reconstructed image and the cophased image in the absence of atmospheric turbulence. The correlation coefficient, $c$ is computed according to the following equation}

\begin{equation}
{ c=\frac{\sum_m\sum_n (A_{mn}-\bar{A})(B_{mn}-\bar{B})}{\sqrt{(\sum_m\sum_n (A_{mn}-\bar{A})^2)(\sum_m\sum_n (B_{mn}-\bar{B})^2)}}}
\end{equation}

{where $A$ and $B$ are the cophased and reconstructed images respectively with size ${m}\times{n} $. The corrosponding correlation coefficients of the reconstructions are given in Table} \ref{tab2}. 

{As it is visible from the reconstructed images, the tomographic speckle imaging algorithm has been found to obtain good reconstructions of the object. But since, the reconstructions rely
on 1-dimensional projections from the image and the reconstructed projections are used to retrieve the object image, the resulting images have artifacts due to the projection operations which results in a lower $c$ value. But when we increase the number of
projections used in radon transformation of image, the quality of reconstruction increases and there are visibly less projection artifacts in recovered image which results in a better $c$ value. The reconstruction quality is still not comparable to the direct bispectrum algorithm as seen from the results and the $c$ values given in the Table} \ref{tab2}. But
tomographic speckle imaging offers a very high computational advantage over direct bispectrum imaging. With lesser number of projections the computational time and memory requirements reduces very significantly. Thus it required to find a trade off between computational time and recovery requirements for larger sized  images and more number of frames.

%

\begin{table}

    \begin{tabular}{ | l | l | l | l |}
    \hline
  
     & Direct Bispectrum & Tomographic Code & Tomographic Code \\ 
     &  & 180 Projections & 18 Projections \\ \cline{2-4}
     Computational Time & 1 hr 21 mins & 15 mins & 2 mins \\ 
     Size Limit for Frames & 200 x 200 pixels & $10^4$ x $10^4$ pixels &  $10^4$ x $10^4$ pixels \\
     {$c$ for 6-star Group}& {0.94} & {0.83} &{0.94} \\
     {$c$ for Extended Object}& {0.96} & {0.92} &{0.95} \\\hline
    
    \end{tabular}
`

\caption{Comparison of computational requirements and reconstruction quality parameter, $c$ of Direct Bispectrum and Tomographic Algorithms. The values are calculated based on MATLAB 32-bit used in a 32-bit Windows system with 4GB of RAM.}
\label{tab2}
\end{table}

\label{sec:2}
\section{Laboratory Simulations}
\begin{figure}
\center
  \includegraphics[width=90mm]{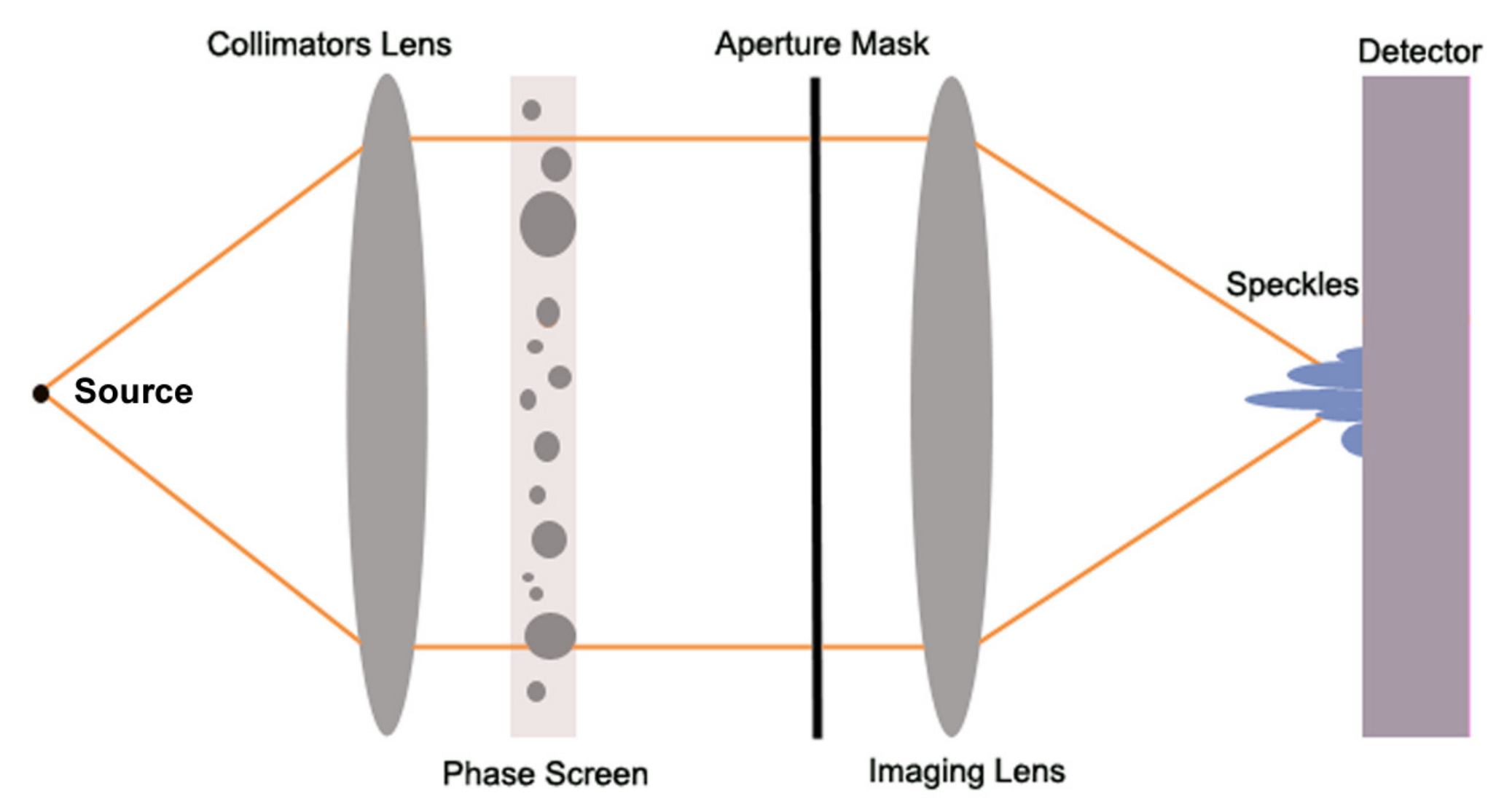}
\caption{Laboratory setup for simulation of aperture masking in presence of atmospheric turbulence}
\label{fig:5}       
\end{figure}

\begin{figure}
\center
  \includegraphics[width=90mm]{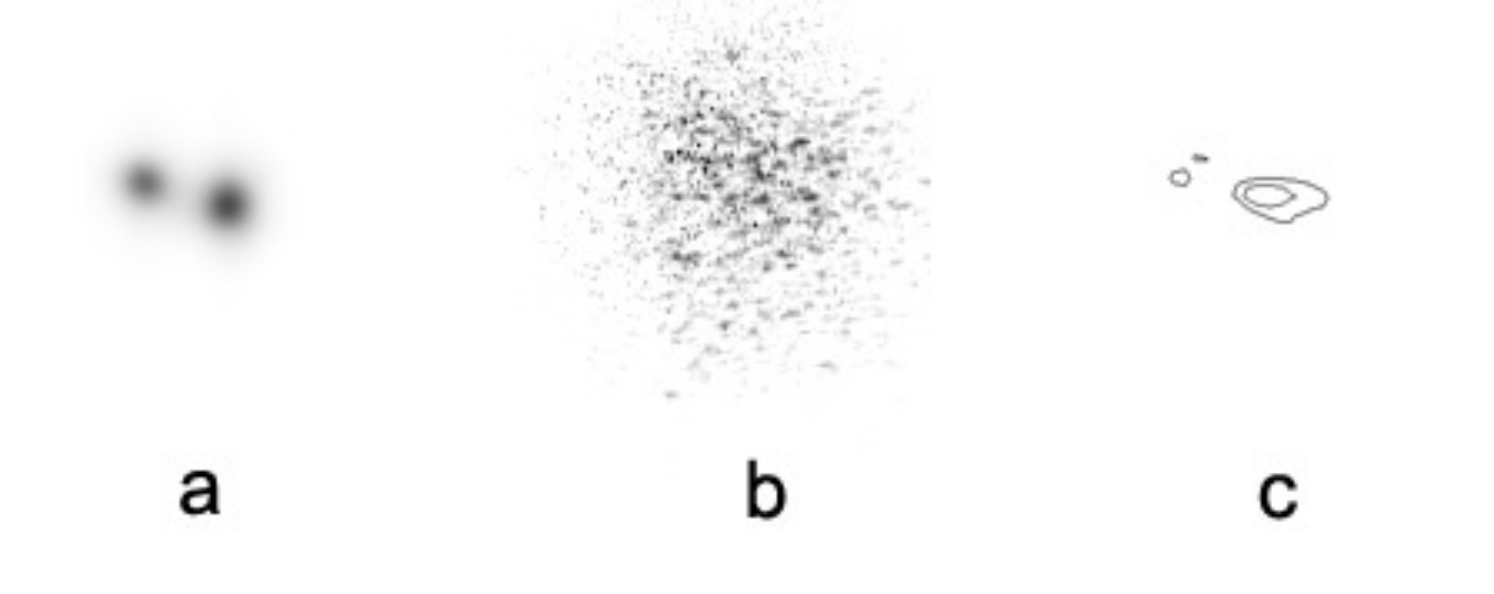}
\caption{Reconstruction from laboratory simulations of speckle imaging. a) Image of the artificial binary star from the experiment with out using turbulence phase screen. b) Speckle image of binary star imaged with a kolmogrov phase screen in laboratory. c) Reconstructed image from Tomographic speckle masking from 100 speckle frames.   }

\label{fig:6}       
\end{figure}

{The experimental setup for the simulation of aperture masking is shown in figure} \ref{fig:5}. {A binary star was simulated using two 5mw laser sourcess which were collimated using a lens. The colllimating lens was followed by a phasescreen, an aperture mask and an imaging lens which focusses on the detector. The laser sources used are 5mW red lasers at 633 nm. The lasers were arranged such that they subtend an angle of 20 arc second in the detector plane. The phase screen is made from spraying glycerine over a glass surface. Several phase screens with different atmospheric coherence lengths were prepared as part of the experiment. The results of the reconstructions using TSM algorithm on the speckle frames of the artificial binary star are shown in figure} \ref{fig:6}. {The speckle images obtained from the experiment had an average of $1000$ counts/frame and $100$ such frames were processed by the Tomographic Speckle Masking code to obtain the reconstructed image. One of the speckle image is shown 
in figure.6.b. 
The results of the reconstructions using TSM algorithm on the speckle frames of the artificial binary star are shown in figure 6.c. The reconstruction resolved the binary stars with the correct seperation of 20 arc seconds. This laboratory experiment clearly demonstrates the performance of tomographic algorithm with aperture masking.} 

\section{Results with Real  Speckle Images} 

\begin{figure}
\center
  \includegraphics[width=90mm]{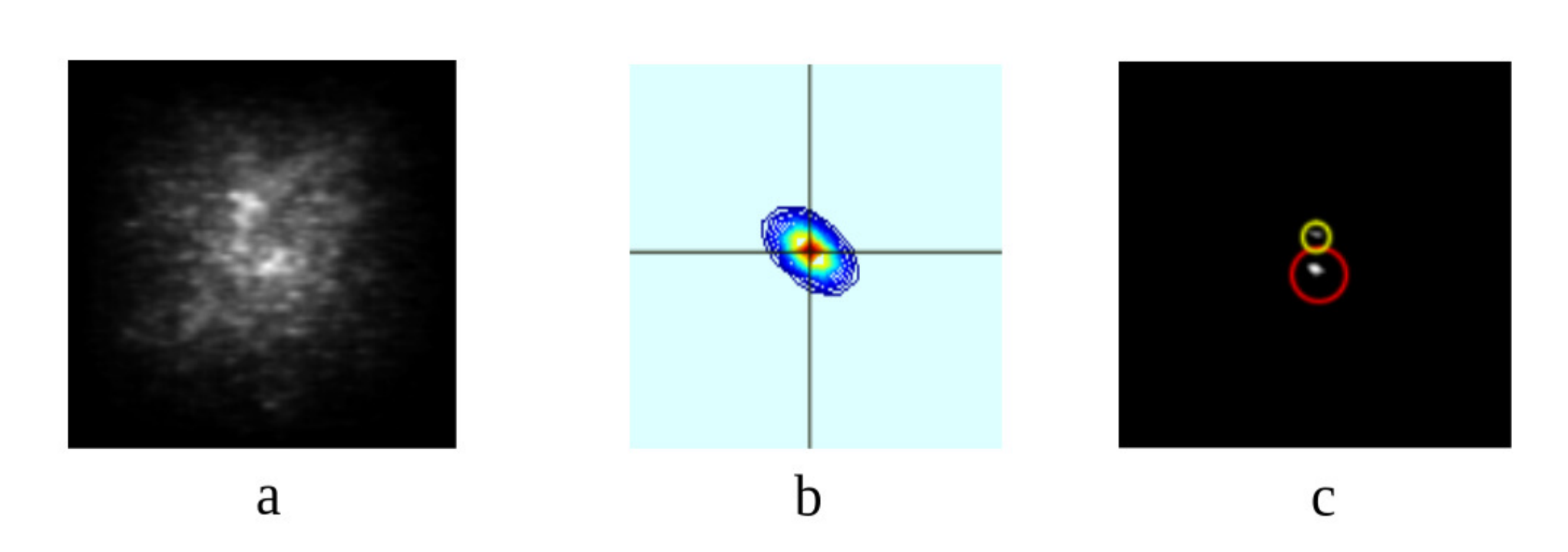}
\caption{Tomographic image reconstruction of speckle images of  $\beta$ Coronae Borealis (HR5747). a) One of the speckle images of HR5747. b) The bispectrum of a one dimensional projection of the speckle image. c) The reconstructed image using TSM Algorithm.}

\label{fig:76}       
\end{figure}
The developed algorithm was applied on images of $\beta$ Coronae Borealis (HR5747) taken on 16 and 17 March 1990 from the 2.34 meter Vainu Bappu Telescope (VBT) in Kavalur using a speckle camera system developed by Saha et al. (1987) \cite{sks5}, the description of which is given below. VBT has two accessible foci for backend instrumentation, such as a prime focus (f/3.25 beam) and a cassegranian focus (f/13 beam). The latter has an image scale of 6.7 arcseconds per mm, which was further slowed down to $\sim$1.21 arcseconds per mm, using a Barlow lens arrangement (Chinnappan et al.,1991 \cite{chin}, Saha et al., 1999 \cite{sks6},). This enlarged image was recorded through a 5~nm filter centred on Hα using an EEV uncooled intensified CCD (ICCD) camera with exposure times of 20~ms. The interface between the 
intensifier screen and the CCD chip is a fibre-optic bundle which reduces the image size by a factor of 1.69. A Data Translation{$^{TM}$ frame-grabber card DT-2851 digitises the video signal. This digitiser resamples the pixels of each row (385 CCD columns to 512 digitized samples) and introduces a net reduction in the row direction of a factor of 1.27. The video frame grabber cards digitize and store the images in the memory buffer of the card. The available frame grabber could store upto two interlaced frames images on its onboard memory. These images are then written onto the hard disc of a personal computer. The observing conditions were fair with an average seeing of ∼2 arcseconds during the nights of 16/17 March 1990. The binary star was earlier resolved using Blind Iterative Deconvolution technique(BID) (Saha \& Venkatakrishnan, 1999 \cite{sks7}) and the seperation was found to be 0.20 arcsecond and magnitude difference of 1.65. The results of tomographic speckle masking on the same speckle images 
gave results show a 
separation of 0.21 arcseconds. The position angles and separations of
the binary components were seen to be consistent with results of the auto-correlation technique and with the published observations of the binary orbit of HR5747 \cite{lab74}. The difference of seperation values with BID and TSM may be attributed to the fact that tomographic speckle masking utilized 8 frames of the binary star as compared to a single frame used in blind iterative deconvolution technique.

\section{Conclusions}

We have successfully developed two algorithms based on triple correlation technique to process atmospherically degraded astronomical images. One uses the direct bispectrum technique and the other is the tomographic speckle masking algorithm that uses radon transformation.
The developed tomographic speckle masking algorithm has been found to obtain good reconstructions with smaller evaluation time and memory requirements, compared to the direct bispectrum technique.
Since the number of projections used for the radon transform can be varied, it is possible to obtain smaller evaluation times at the cost of reconstruction
quality.
The developed algorithm can be used with short exposure speckle frames to obtain astronomical images of high resolution. 
Bayesian and Regularization techniques could be used to improve the reconstructions from the algorithm and also enable it to be used with long baseline interferometric data.
Tomographic speckle imaging offers a unique way to implement very fast reconstruction algorithms based on triple correlation. It has been shown with the help of simulations, laboratory experiments and real observations
that tomographic speckle masking code can be used with aperture masking interferometry to obtain very good reconstructions of stellar objects.


\end{document}